\documentclass[12pt]{article}
\usepackage{latexsym,amsfonts,amssymb}
\makeatletter
\@addtoreset{equation}{subsection}
\makeatother

\begin{document}
\begin{center}
{\Large \bf Covariance, Geometricity, Setting, and Dynamical
Structures on Cosmological Manifold}
\\[1.5cm]
 {\bf Vladimir S.~MASHKEVICH}\footnote {E-mail:
  Vladimir\_Mashkevich@qc.edu}
\\[1.4cm] {\it Physics Department
 \\ Queens College\\ The City University of New York\\
 65-30 Kissena Boulevard\\ Flushing, New York
 11367-1519} \\[1.4cm] \vskip 1cm

{\large \bf Abstract}
\end{center}

The treatment of the principle of general covariance based on
coordinate systems, i.e., on classical tensor analysis suffers
from an ambiguity. A more preferable formulation of the principle
is based on modern differential geometry: the formulation is
coordinate-free. Then the principle may be called ``principle of
geometricity.'' In relation to coordinate transformations, there
had been confusions around such concepts as symmetry, covariance,
invariance, and gauge transformations. Clarity has been achieved
on the basis of a group-theoretical approach and the distinction
between absolute and dynamical objects. In this paper, we start
from arguments based on structures on cosmological manifold rather
than from group-theoretical ones, and introduce the notion of
setting elements. The latter create a scene on which dynamics is
performed. The characteristics of the scene and dynamical
structures on it are considered.

\newpage

\section*{Introduction}

In a fundamental work on the general theory of relativity [1],
Einstein gave due attention to the principle of general covariance
as one of the cornerstones of the theory. The principle was
formulated as the requirement that the general laws of nature must
be expressed in terms of equations valid in all coordinate
systems. However, Kretschmann [2] argued that equations originally
written in any coordinate system may be extended to all coordinate
systems and thus made covariant; therefore the principle of
general covariance involves no physical content. Einstein
concurred with the argumentation [3].

The treatment of the principle of general covariance based on
coordinate systems, i.e., on classical tensor analysis, as will be
seen later, suffers from an ambiguity---as long as the geometric
character of quantities is not specified in advance. A more
preferable formulation of the principle is based on modern
differential geometry: such a formulation is coordinate-free. We
quote [4]: ``Every physical quantity must be describable by a
(coordinate-free) geometric object, and the laws of physics must
all be expressible as geometric relationships between these
geometric objects.'' In such a formulation, the principle may be
called ``principle of geometricity.''

In relation to coordinate transformations, there were confusions
around such concepts as symmetry, covariance, invariance, and
gauge transformation [5]. In a textbook, the point was for the
first time made clear by Anderson [6]. His treatment is based on a
group-theoretical approach and the distinction between absolute
and dynamical objects (see also [7], [5]).

In this paper, we start from arguments based on structures on
cosmological manifold rather than from group-theoretical ones.
Therefore we introduce the notion of setting objects---instead of
absolute ones. The setting objects create a scene on which
dynamics is performed.

The purpose of the paper is to consider briefly the
characteristics of the scene and dynamical structures on it.

\section{Covariance and geometricity}

\subsection{Cosmological manifold}

A primary setting object, or element is a cosmological manifold,
i.e., a smooth 4-manifold [8] $$M=M^{4}\,,\quad p\in M $$ At this
juncture, there is no structure on $M$.

\subsection{The principle of general covariance in terms
of coordinate \\systems}

The principle of general covariance was originally formulated on
the basis of classical tensor analysis, in which it is necessary
to exploit coordinate systems. In this approach, tensor quantities
are defined in terms of their components and of the transformation
rules for the latter under coordinate changes. The principle is
formulated as follows [1]:

{\it ``The general laws of nature are to be expressed by equations
which hold good for all systems of co-ordinates, that is, are
co-variant with respect to any substitutions whatever (generally
co-variant).''}

Kretschmann argued that any equation written in an arbitrary
coordinate system may be rewritten in any other coordinate
system---on the basis of the transformation rules.

\subsection{Ambiguity}

There is an ambiguity in the application of the transformation
rules---as long as the tensor character of quantities involved in
the equation is not specified. This is an example. Let four
equations be given in a coordinate system
$x=(x^{\mu})_{\mu=0}^{3}$:

 $$ f_{1}^{(\nu)}(x)=f_{2}^{(\nu)}(x)$$
where $\nu=0,1,2,3$ is the label of the $f$. Let
$\bar{x}=(\bar{x}^{\mu})$ be another coordinate system:

$$M\ni p\leftrightarrow x\leftrightarrow \bar{x}$$ There are
different possibilities:

1) both $f_{1}^{(\nu)}$ and $f_{2}^{(\nu)}$ are functions (i.e.,
scalars); then

$$f_{n}^{(\nu)}(x)=f_{n}^{(\nu)}(x(\bar{x}))=:\bar{f}_{n}^{(\nu)}(\bar{x})$$
and we have the implication

$$f_{1}^{(\nu)}(x)=f_{2}^{(\nu)}(x)\Rightarrow
\bar{f}_{1}^{(\nu)}(\bar{x})=\bar{f}_{2}^{(\nu)}(\bar{x})$$ or

$$f_{1}^{(\nu)}(p)=f_{2}^{(\nu)}(p)\,,\quad p\in M$$ which is
covariant.

2) both $f_{1}^{(\nu)}$ and $f_{2}^{(\nu)}$ are components of
vectors:

$$f_{n}^{(\nu)}=v_{n}^{\nu}$$ then
$f_{1}^{(\nu)}(x)=f_{2}^{(\nu)}(x)$ amounts to

$$v_{1}^{\nu}=v_{2}^{\nu}$$ which is fulfilled in all coordinate
systems, i.e., is covariant.

3) $f_{1}^{(\nu)}$ represents a vector, whereas $f_{2}^{(\nu)}$ is
a function:

$$f_{1}^{(\nu)}=v^{\nu}\,,\quad f_{2}^{(\nu)}=f^{(\nu)}$$ then

$$f_{1}^{(\nu)}(x)=f_{2}^{(\nu)}(x)\nRightarrow
v^{\nu}=f^{(\nu)}$$

\subsection{The principle of geometricity}

To avoid the ambiguity we have to be based on modern differential
geometry rather than on classical tensor analysis, and formulate
the principle of geometricity:

Spacetime structure and spacetime aspects of matter objects must
be expressed in terms of geometric notions.

Now in the above example, the $f_{n}^{(\nu)}$ should be specified
as geometric objects.

\section{Setting}

\subsection{Setting as a scene for performing dynamics}

We consider a physical theory that involves dynamics on
cosmological manifold $M$. (A definition of dynamics will be given
below.) All physical objects are classified into two categories:
setting objects and dynamical ones. A setting is a family of
setting objects; a dynamical system is a set of dynamical objects.
The mathematical representation of the setting is independent of
the dynamical system, whereas the representation of the latter
involves the setting.

Figuratively speaking, a setting is a scene on which dynamics is
performed.

\subsection{Natural and free setting elements}

The setting objects are classified into two subcategories: natural
and free objects. The natural setting is induced by dynamics in
the sense that the latter involves the former. The free setting
elements, if any, play an auxiliary role.

These are the examples of natural setting elements: affine
structures of Aristotelian and Newtonian spacetimes [9], the
Minkowskian metric, the gravitational and cosmological constants,
interaction constants of quantum field theory; and of free setting
elements: reference frames (tetrads), coordinate systems.

\subsection{The principle of minimal (free) setting}

Now we may endow the principle of geometricity with a certain
constructive meaning. In view of that principle, we advance the
principle of minimal (free) setting:

A (free) setting should include as few elements as possible.

It is the absence of free setting elements that is in accordance
with the principle of geometricity.

\section{Objects and related fields}

\subsection{Objects and fields}

Let $w$ be an object which is an element of a set $\mathcal{W}$,

$$w\in \mathcal{W}$$ A related field is defined on a submanifold
[8] of cosmological manifold,

$$M'\subset M$$ as

$$w_{M'}\in \prod_{p\in M'}\mathcal{W}_{p}\,,\;\;w_{M'}(p)\in
\mathcal{W}_{p}$$  $\mathcal{W}_{p}$ is related to $p\in M$.

Introduce an abridged notation:

$$w_{M'}(p)=:w(p)$$

\subsection{Geometric quantities}

Let

$$F_{\gamma}\,,\;\;\gamma\in\Gamma$$ be a geometric quantity,
$\gamma$ being the type of the latter: scalar, vector, tensor,
spinor. $F_{\gamma}$ is an element of a space
$\mathcal{F}_{\gamma}$\,,

$$F_{\gamma}\in \mathcal{F}_{\gamma}$$ Introduce

$$\mathcal{F}_{U}:=\bigcup_{\gamma\in\Gamma}\mathcal{F}_{\gamma}\,,\quad
F\in\mathcal{F}_{U} $$ where $F$ is a generic $F_{\gamma}$.

Geometric fields are defined according to the preceding
subsection.

\subsection{Variables, states, and valuables}

Introduce variables:

$$v_{\gamma b}\,,\quad b\in \mathcal{B}$$ variable value sets:

$$\mathcal{V}_{\gamma}\,,\quad v_{\gamma
b}\in\mathcal{V}_{\gamma}$$
$$\mathcal{V}_{U}:=\bigcup_{\gamma\in\Gamma}\mathcal{V}_{\gamma}\,,\quad
v\in\mathcal{V}_{U} $$ and states:

$$\omega\in\Omega$$ In the final analysis, it is the expectation
values of variables in states that have immediate physical
meaning. Therefore we introduce the notion of valuable:

$$\langle\;\rangle:\mathcal{V}_{U}\times\Omega\rightarrow\mathcal{F}_{U}\,,\quad
(v,\omega)\mapsto\langle v,\omega\rangle\in\mathcal{F}_{U} $$ or,
in more detail

$$\langle v_{\gamma b},\omega\rangle\in\mathcal{F}_{\gamma}$$

\subsection{Classical variables and fields}

Introduce the following notation for classical variables:

$$v_{\gamma b}=\xi_{\gamma b}\in\Xi_{\gamma}\,,\quad
b=b^{\mathrm{class}}\in\mathcal{B}^{\mathrm{class}}$$ In classical
physics, no distinction is usually made between an abstract
variable and its expectation value [10]. So we put

$$\langle\xi_{\gamma
b},\omega^{\mathrm{class}}\rangle=:\xi_{\gamma
b}\in\mathcal{F}_{\gamma}$$ For a classical field, we have a
notation

$$\xi_{\gamma bM'}\,,\quad\xi_{\gamma b}(p)\,,\;p\in M'$$

\subsection{Quantum variables and fields}

Introduce the following designations: the Hilbert space
$\mathcal{H}$\,,

$$\hat{\mathcal{A}}:=L(\mathcal{H,H})\,,\quad
\hat{A}\in\hat{\mathcal{A}}\,,\quad \hat{A}:\mathcal{H\rightarrow
H }$$ A quantum entity (variable or field) generally consists  of
two components: classical $F_{\gamma}$ and properly quantum
$\hat{A}$. Namely, a quantum variable

$$\hat v_{\gamma b}=F_{\gamma b}\hat{A}_{b}:=F_{\gamma
b}\otimes\hat{A}_{b}\,,\quad
b=b^{\mathrm{quant}}\in\mathcal{B}^{\mathrm{quant}}$$ where
$\hat{A}$ as a geometric quantity is considered to be a scalar.
For a valuable we have

$$\langle \hat v_{\gamma
b},\omega^{\mathrm{quant}}\rangle=F_{\gamma
b}\langle\hat{A}_{b},\omega^{\mathrm{quant}}\rangle$$ and (for a
pure state)

$$\langle\hat{A},\omega^{\mathrm{quant}}\rangle=
(\Psi,\hat{A}\Psi)\in\mathbb{C}\,,\quad\Psi\in\mathcal{H}$$ so
that

$$\langle \hat v_{\gamma
b},\omega^{\mathrm{quant}}\rangle\in\mathcal{F}_{\gamma}$$
Generally

$$\hat v_{\gamma}\in\mathcal{F}_{\gamma}\otimes\hat{\mathcal{A}}$$
or

$$\hat v_{\gamma b}=\int\limits_{\mathcal{L}}\mu(dl)F_{\gamma
bl}\hat{A}_{bl}$$ and a valuable

$$\langle \hat v_{\gamma
b},\omega^{\mathrm{quant}}\rangle=\int\limits_{\mathcal{L}}\mu(dl)F_{\gamma
bl}(\Psi,\hat{A}_{bl}\Psi)$$

A quantum field $\hat v_{\gamma bM'}$ may be described as follows:

$$\hat v_{\gamma
b}(p)=\int\limits_{\mathcal{L}(p)}\mu_{p}(dl)F_{\gamma bl
}\hat{A}_{bl}$$

$$\langle \hat v_{\gamma
b}(p),\omega^{\mathrm{quant}}\rangle=\int\limits_{\mathcal{L}(p)}\mu_{p}(dl)F_{\gamma
bl }(\Psi,\hat{A}_{bl}\Psi)$$

\section{Dynamics on cosmological manifold without structure}

\subsection{Dynamics}

Dynamics on $M'\subset M$ is a family of valuable fields:

$$\{\xi_{\gamma
bM'}:\gamma\in\Gamma\,,\;b\in\mathcal{B}^{\mathrm{class}}\}$$ and

$$\{\langle \hat v_{\gamma
b}(p),\omega^{\mathrm{quant}}\rangle:\gamma\in\Gamma\,,\;
b\in\mathcal{B}^{\mathrm{quant}}\,,\;p\in M'\}$$ Classical
dynamics is constructed on the basis of the $\xi$ themselves,
quantum dynamics is constructed on the basis of the
$\mathcal{F}_{U}\,,\;\mathcal{H}$, and $\hat{\mathcal{A}}$.

\subsection{Mode-series expansion: Manifold modes}

Let us introduce the expansion of a quantum field in terms of
manifold modes:

$$\hat v_{\gamma b}(p)=\int\limits_{\mathcal{M}}\mu(dm)
\sum_{n\in\mathcal{N}_{\gamma}}F_{\gamma bmn }(p)\hat{A}_{bmn}$$

$$\langle \hat v_{\gamma
b},\omega^{\mathrm{quant}}\rangle=\int\limits_{\mathcal{M}}\mu(dm)
\sum_{n\in\mathcal{N}_{\gamma}}F_{\gamma
bmn}(p)(\Psi,\hat{A}_{bmn}\Psi)$$ The set

$$\{F_{\gamma
mnM'}:m\in\mathcal{M}\,,\;n\in\mathcal{N}_{\gamma}\}$$ of manifold
modes forms a complete system on $M'$.

Now we put

$$F_{\gamma mn}(p)=f_{m}(p)e_{\gamma mn}(p)$$ where $f_{mM'}$ is a
scalar field on $M'$. The set

$$\{f_{mM'}:m\in\mathcal{M}\}$$ forms a complete system on $M'$,
and the set

$$\{e_{\gamma mn}(p):n\in\mathcal{N}_{\gamma}\}$$ forms a complete
system at $p\in M'$.

\section{The Cauchy problem and manifold foliation}

\subsection{A Cauchy surface and a foliation}

Let $M$ possess a Cauchy surface. Then there exists a foliation of
$M$ [11], [12]:

$$M=T\times S\,,\quad M\ni p=(t,s)\,,\;t\in T\,,\;s\in S$$ where
1-manifold $T$ is a cosmological time and 3-manifold $S$ is a
cosmological space. The tangent space $M_{p}$ at a point $p\in M$
is

$$M_{p}=T_{t}\oplus S_{s}\,,\quad p=(t,s)$$ A Cauchy surface

$$M_{\mathrm{C}}=\{t_{0}\}\times S\ni p=(t_{0}\,,s)$$ specifies a
unique foliation (by means of synchronous coordinates [13]). In
the synchronous reference (i.e., in every synchronous reference
frame)

$$T_{t}\perp S_{s}$$

As to the choice of a Cauchy surface, notice the following. If
metric is given, different surfaces generally give rise to
different foliations; however, if the Cauchy problem includes the
determination of metric, the choice of the surface in general does
not affect physical results.

Thus as long as dynamics is constructed starting from initial
conditions, a natural construction involves a Cauchy surface with
the associated foliation and synchronous reference.

Now

$$M'=T'\,,\quad T'\subset T$$

\subsection{Initial conditions}

Initial conditions for classical fields are of the form

$$\{(\xi\,,\;\partial_{t}\xi)_{M_{\mathrm{C}}}\}$$ which
corresponds to second order dynamics.

For quantum fields we have

$$\{\hat v_{M_{\mathrm{C}}}\;\;\mathrm{or}\;\; (\hat
v\,,\;\partial_{t}\hat v)_{M_{\mathrm{C}}}\}$$ or

$$\{F_{mnM_{\mathrm{C}}}\;\;\mathrm{or}\;\;(F_{mn}\,,
\partial_{t}F_{mn})_{M_{\mathrm{C}}}\}$$
which corresponds to first or second order dynamics, respectively.

\section{Dynamics on a foliated manifold}

\subsection{Time dependent quantum objects}

As long as cosmological manifold is foliated, it is natural to
introduce time dependent quantum objects:

operator

$$\hat{A}_{T'}\in\hat{\mathcal{A}}^{T'}\,,\quad
\hat{A}_{T'}(t)=:\hat{A}(t)\in\hat{\mathcal{A}}$$

state

$$\omega_{T'}\in\Omega^{T'}\,,\quad\omega(t)\in\Omega\,,
\quad\Psi_{T'}\in\mathcal{H}^{T'}\,,\quad\Psi(t)\in\mathcal{H}$$

valuable

$$(\Psi(t),\hat{A}(t)\Psi(t))$$

\subsection{Mode-series expansion: Space modes}

We introduce the expansion of a quantum field in terms of space
modes:

$$\hat v_{\gamma b\{t\}\times
S}=\int\limits_{\mathcal{M}}\mu(dm)\sum_{n\in\mathcal{N_{\gamma}}}F_{\gamma
bmn\{t\}\times S }\hat{A}_{bmn}(t)\,,\quad
b\in\mathcal{B}^{\mathrm{quant}}$$ or

$$\hat v_{\gamma
b}(t,s)=\int\limits_{\mathcal{M}}\mu(dm)\sum_{n\in\mathcal{N}_{\gamma}}F_{\gamma
bmn }(t,s)\hat{A}_{bmn}(t)$$ so that

$$\langle \hat v_{\gamma
b}(t,s),\omega^{\mathrm{quant}}\rangle=\int\limits_{\mathcal{M}}\mu(dm)
\sum_{n\in\mathcal{N}_{\gamma}}F_{\gamma
bmn}(t,s)(\Psi(t),\hat{A}_{bmn}(t)\Psi(t))$$ Next we put

$$F_{\gamma mn}(t,s)=f_{m}(t,s)e_{\gamma mn}(t,s)$$ The $F_{\gamma
mn\{t\}\times S }$ and $f_{m\{t\}\times S}$ are time dependent
space modes.

The sets

$$\{F_{\gamma mn\{t\}\times
S}:m\in\mathcal{M}\,,\;n\in\mathcal{N_{\gamma}}\}$$ and

$$\{f_{m\{t\}\times S}:m\in\mathcal{M}\}$$ form complete systems
on $\{t\}\times S$, the set

$$\{e_{\gamma mn}(t.s):n\in\mathcal{N}_{\gamma}\}$$ forms a
complete system at $(t,s)$.

Now the initial conditions are

$$(\{\hat{A}_{bmn}(t_{0}):b\in\mathcal{B}^{\mathrm{quant}}\,,
\;m\in\mathcal{M}\,,\;n\in\mathcal{N}\}\,,\;\;\Psi(t_{0}))$$ which
corresponds to first order dynamics.

\subsection{Dynamical pictures}

There are these dynamical pictures:

the Schr\"odinger picture:

$$\Psi_{S}=U_{S}(t,t_{0})\Psi(t_{0})\,,\quad\hat{A}_{S}=\mathrm{const}$$

the Heisenberg picture:

$$\Psi_{H}=\Psi_{S}(t_{0})=\mathrm{const}\,,
\quad\hat{A}_{H}(t)=U_{S}^{\dag}(t,t_{0})\hat{A}_{S}U_{S}(t,t_{0})$$

a generic picture:

$$\Psi(t)=U_{1}(t)\Psi_{S}(t_{0})\,,\quad\hat{A}(t)=
U_{2}^{\dag}(t)\hat{A}_{S}U_{2}(t)\,,\quad
U_{2}(t)U_{1}(t)=U_{S}(t,t_{0}) $$

Note that a Schr\"odinger variable $\hat v_{S}$ depends on $t$
through $F$.

\section{Setting elements}

\subsection{Manifold}

Let us list setting elements in the above structures.

Cosmological manifold $M^{4}$ is a primary natural setting element
involved in all structures.

\subsection{Initial conditions and Cauchy surface}

Dynamics implies initial conditions, and the latter involve a
Cauchy surface. So initial conditions and a Cauchy surface are
natural setting elements.

\subsection{Foliation}

In general, a foliation is not unique. So let

$$ M=T\times
S\;\;\mathrm{and}\;\,M=\overline{T}\times\overline{S}\,,\quad M\ni
p\leftrightarrow(t,s)\leftrightarrow(\bar{t},\bar{s})$$ Then there
are modes

$$f_{m\{t\}\times S}=:f_{mt}(s)$$ and

$$\bar{f}_{\bar{m}\{\bar{t}\}\times\overline{S}}=:
\bar{f}_{\bar{m}\bar{t}}(\bar{s})$$ Let

$$\varphi(t,s)=\int\limits_{\mathcal{M}}\mu(dm)c_{m}(t)f_{mt}(s)$$
We have

$$\bar{\varphi}(\bar{t},\bar{s})=\bar{\varphi}(\bar{t}(t,s),\bar{s}(t,s))
=:\tilde{\bar{\varphi}}(t,s)=
\int\limits_{\mathcal{M}}\mu(dm)\tilde{\bar{c}}_{m}(t)f_{mt}(s) $$
Thus

$$\bar{f}_{\bar{m}\bar{t}}(\bar{s})=\int\limits_{\mathcal{M}}\mu(dm)
\tilde{\bar{c}}_{\bar{m}m}(t)f_{mt}(s)$$ The
$\tilde{\bar{c}}_{\bar{m}m}(t)$ are functions of $t$, so that
different foliations are not equivalent, and generally a foliation
is a free setting element. But as long as a Cauchy surface is
specified, the related foliation is a natural setting element.

\subsection{Setting for manifold and space modes}

The setting for manifold modes is a choice of them and initial
conditions for them. The setting is free.

The setting for space modes is a foliation $M=T\times S$ and
initial conditions for $\hat{A}_{mn}$ and $\Psi$. As long as a
Cauchy surface is specified, the setting is natural.

\section*{Acknowledgments}

I would like to thank Alex A. Lisyansky for support and Stefan V.
Mashkevich for helpful discussions.

\end{document}